\documentclass[twocolumn,aps,prb,superscriptaddress,citeautoscript]{revtex4}

\usepackage{graphicx}
\usepackage{dcolumn}
\usepackage{bm}
\usepackage{amsmath}
\usepackage[T1]{fontenc}
\usepackage{sidecap}
\usepackage{float}
\usepackage{siunitx}
\begin{document}

\preprint{}

\title{Field-free spin-orbit torque switching in Co/Pt/Co multilayer with mixed magnetic anisotropies}

\author{Stanis\l{}aw \L{}azarski}
 \email{lazarski@agh.edu.pl}
\author{Witold Skowro\'{n}ski}
 \email{skowron@agh.edu.pl}
 \author{Jaros\l{}aw Kanak}
\affiliation{AGH University of Science and Technology, Department of Electronics, Al. Mickiewicza 30, 30-059 Krak\'{o}w, Poland}
\author{\L{}ukasz Karwacki}
\affiliation{AGH University of Science and Technology, Department of Electronics, Al. Mickiewicza 30, 30-059 Krak\'{o}w, Poland}
\affiliation{Institute of Molecular Physics, Polish Academy of Sciences, ul. Smoluchowskiego 17, 60-179 Pozna\'{n}, Poland}
\author{S\l{}awomir Zi\k{e}tek}
\affiliation{AGH University of Science and Technology, Department of Electronics, Al. Mickiewicza 30, 30-059 Krak\'{o}w, Poland}
\author{Krzysztof Grochot}
\affiliation{AGH University of Science and Technology, Department of Electronics, Al. Mickiewicza 30, 30-059 Krak\'{o}w, Poland}
\affiliation{AGH University of Science and Technology, Faculty of Physics and Applied Computer Science, Al. Mickiewicza 30, 30-059 Krak\'{o}w, Poland}
\author{Tomasz Stobiecki}
\affiliation{AGH University of Science and Technology, Department of Electronics, Al. Mickiewicza 30, 30-059 Krak\'{o}w, Poland}
\affiliation{AGH University of Science and Technology, Faculty of Physics and Applied Computer Science, Al. Mickiewicza 30, 30-059 Krak\'{o}w, Poland}
\author{Feliks Stobiecki}
\affiliation{Institute of Molecular Physics, Polish Academy of Sciences, ul. Smoluchowskiego 17, 60-179 Pozna\'{n}, Poland}

\date{\today}

\begin{abstract}
Spin-orbit-torque (SOT) induced magnetization switching in Co/Pt/Co trilayer, with two Co layers exhibiting magnetization easy axes orthogonal to each other is investigated. Pt layer is used as a source of spin-polarized current as it is characterized by relatively high spin-orbit coupling. The spin Hall angle of Pt, $\theta=0.08$ is quantitatively determined using spin-orbit torque ferromagnetic resonance technique. In addition, Pt serves as a spacer between two Co layers and depending on it's thickness, different interlayer exchange coupling (IEC) energy between ferromagnets is induced.  Intermediate IEC energies, resulting in a top Co magnetization tilted from the perpendicular direction, allows for SOT-induced field-free switching of the top Co layer. The switching process is discussed in more detail, showing the potential of the system for neuromorphic applications.
\end{abstract}

\maketitle

\section{Introduction}
Manipulation of the magnetization of micro- and nano-structures by electrical means allows for a design of low energy and scalable spintronic devices~\cite{bhatti_2017}. A number of applications taking advantage of the magnetic tunnel junctions (MTJs) and spin-valve structures have been proposed, that utilize magnetoresistance and spin-transfer torque (STT)~\cite{slonczewski_1996,berger_1996} effects for operation. Magnetic tunnel junctions with in-plane magnetic easy axis based on MgO have been among the most studied devices due to high tunnelling magnetoresistance (TMR) ratio~\cite{ikeda_2008}, and current-induced magnetization switching~\cite{huai_2004} 
observed in the absence of an external magnetic field. Unfortunately, in general they require high critical current densities for the magnetization switching. Recently, MTJs with perpendicular easy axis have been proposed to reduce this switching current and to enhance thermal stability of the free layer~\cite{ikeda_2010,nakayama_2008,amiri_2011}. Nonetheless, the current densities used for switching may degrade MgO tunnel barrier properties with time, which hampers potential endurance of the device. On the other hand, spin-orbit torque (SOT) driven switching of magnetization with perpendicular magnetic anisotropy (PMA) has emerged and gained a lot of attention as it does not require high density current flow via thin MgO tunnel barrier. Initially SOT switching of perpendicularly magnetized ferromagnets has relied on an external magnetic field to break the time-reversal symmetry~\cite{miron_2011,liu_2012PRL,liu_2012}. However, it would be impractical to use external field for devices application. By using antiferromagnets, the symmetry can be broken and magnetization switching in the absence of the external magnetic field can be acheived.~\cite{mendes_2014,ou_2016,oh_2016,fukami_2016,wu_2016,razavi_2017} \newline
In this work we present another concept utilizing two Co layers with orthogonal easy axes\cite{balaz_2009,lau_2016} that are coupled via interlayer exchange coupling (IEC) across Pt spacer - as for example in Ref.~[\onlinecite{matczak_2015}], to achieve SOT switching without an external magnetic field.\cite{yu_2014,you_2015} The magnetization of the top Co layer is tilted from the perpendicular direction due to ferromagnetic coupling with an in-plane magnetized bottom Co layer. In addition, the Pt spacer serves as a source of the spin current due to spin-orbit coupling characterised by moderate spin Hall angle ($\theta_{\text{SHE}}$).\cite{liu_2012PRL,emori_2013,woo_2014} Moreover, non-orthogonal direction of Co layer magnetizations results in analogue-like switching behaviour. This property of the device is potentially useful as a building block of neuromorphic network utilizing spintronic elements~\cite{fukami_2016}.

\section{Experiment}
\begin{figure}[H]
\centering
\includegraphics[width=\columnwidth]{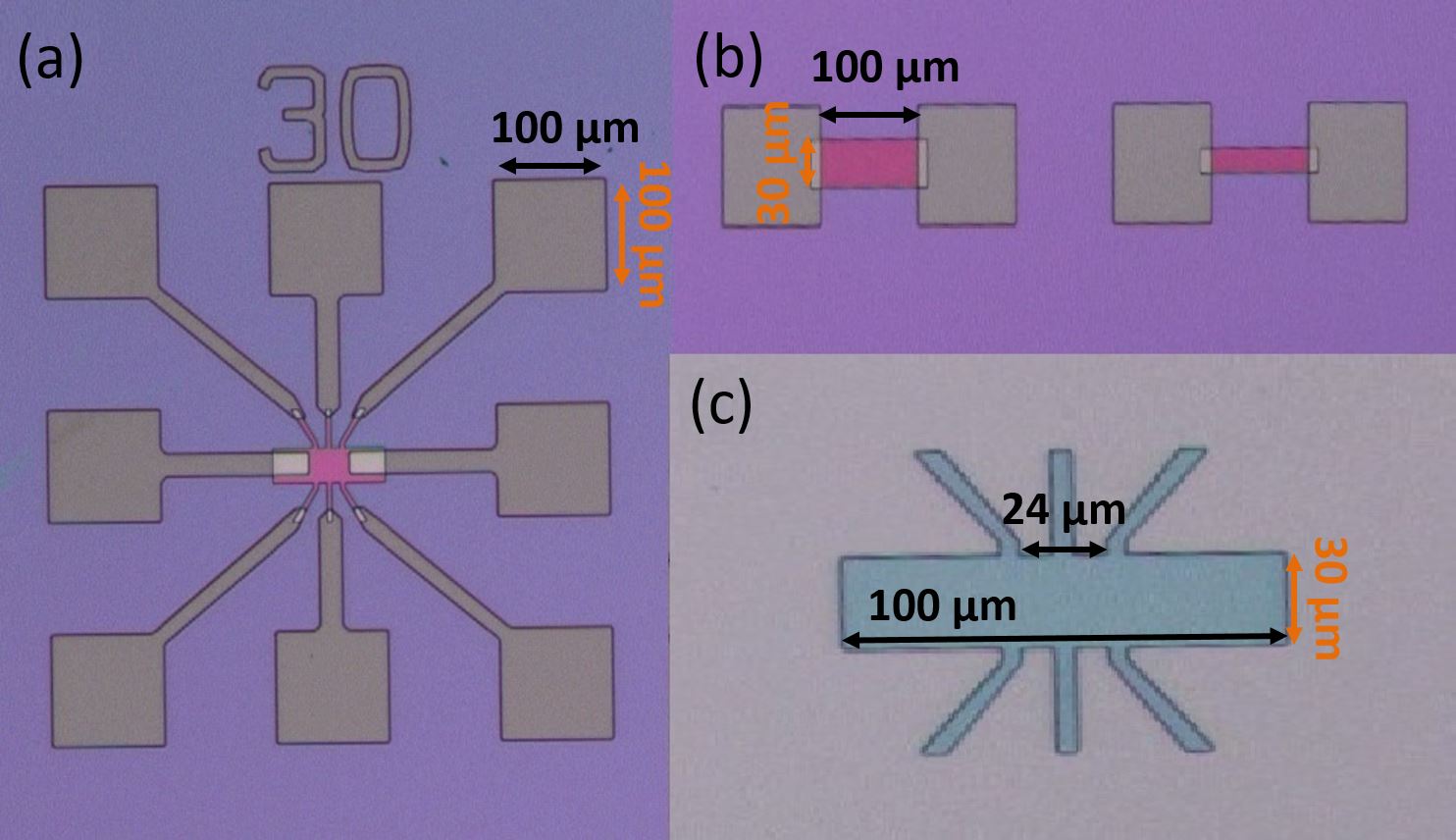}
\caption{Layout of the fabricated devices used for different measurements: (a) Hall bar for SOT-induced switching, (b) stripes of different width for SOT-dynamics. Detailed dimensions of the Hall bar are given in (c).}
\label{fig:spider}
\end{figure}

The following multilayer structure is investigated: Si/SiO$_{2}$/Ti(2)/Co(3)/Pt($t_\mathrm{Pt}$)/Co(1)/MgO(2)/Ti(2) (thickness in nm) with $t_\mathrm{Pt}$ varying from 0 to 4 nm. The thicker Co layer exhibits in-plane anisotropy, whereas the thinner one is characterized by an effective perpendicular anisotropy. In order to determine the resistivity and SOT dynamics of Pt and Co, additional multilayer structures of Si/SiO$_{2}$/Co(5)/Pt(0-10) and Si/SiO$_{2}$/Ti(2)/Co(0-10)/Pt(4) were prepared. 
All samples were deposited using magnetron sputtering system in Ar atmosphere. After the deposition process, all samples were characterized by X-ray diffraction (XRD). X-ray reflectivity (XRR) measurements was made in order to control the thickness of particular layers in the system. The texture was inspected using rocking curve and polar figure methods (See Supplemental Material\cite{supplemental_material}). The Hall bars and stripes dedicated for spin-orbit torque ferromagnetic resonance (SOT-FMR)  measurements were fabricated using electron beam lithography, ion-beam etching and lift-off process.  
The nominal dimensions of samples were: $w\times l$ where  $w=10, 20, 30~\mu\mathrm{m}$ and $l=100~\mu\mathrm{m}$ for SOT-induced dynamics measurements and Hall bars of $w=10, 20, 30~\mu$m and $l=24~\mu$m for current induced magnetization switching experiments - Fig. \ref{fig:spider}. Electrical connections were fabricated using Al(20)/Au(30) contact pads of 100 $\times$ 100 $\mu$m. A dedicated multi-probe system, which enables sample rotation in a given plane under external magnetic field, was used.

The resistivity of Co and Pt layers were determined in Pt and Co wedge sample, as described in Ref.~[\onlinecite{kawaguchi_2018}]. The Pt resistivity was almost constant $\rho_\mathrm{Pt} \approx$ 23 $\mu\Omega$cm whereas, the Co resistivity, $\rho_{\text{Co}}$, varied from 15 to 29 $\mu\Omega$cm with decreasing thickness - detailed information are presented in Supplemental Material\cite{supplemental_material}. We note that  $\rho_\mathrm{Pt}$ is smaller than in our previous work~\cite{skowronski_2018} because in this case Pt was deposited on high textured Co in contrast to Pt deposited on amorphous CoFeB. The SOT-FMR measurements were conducted in an in-plane magnetic field applied at an 45$^\circ$ angle with respect to the long stripe axis and RF signal power of 16 dBm.
The DC voltage originating from the mixing between oscillating resistance (due to SOT and magnetoresistance) and in-phase charge current was measured using lock-in amplifier, which was synchronized with an amplitude modulated RF source. The SOT-induced switching experiment was performed in both in-plane and perpendicular magnetic fields. 
In order to show the analogue-like behaviour, the experiment with varying time of the switching pulses between 500 $\mu$s and 500 ms was performed.

\section{Results and discussion}
\begin{figure}[H]
\centering
\includegraphics[width=\columnwidth]{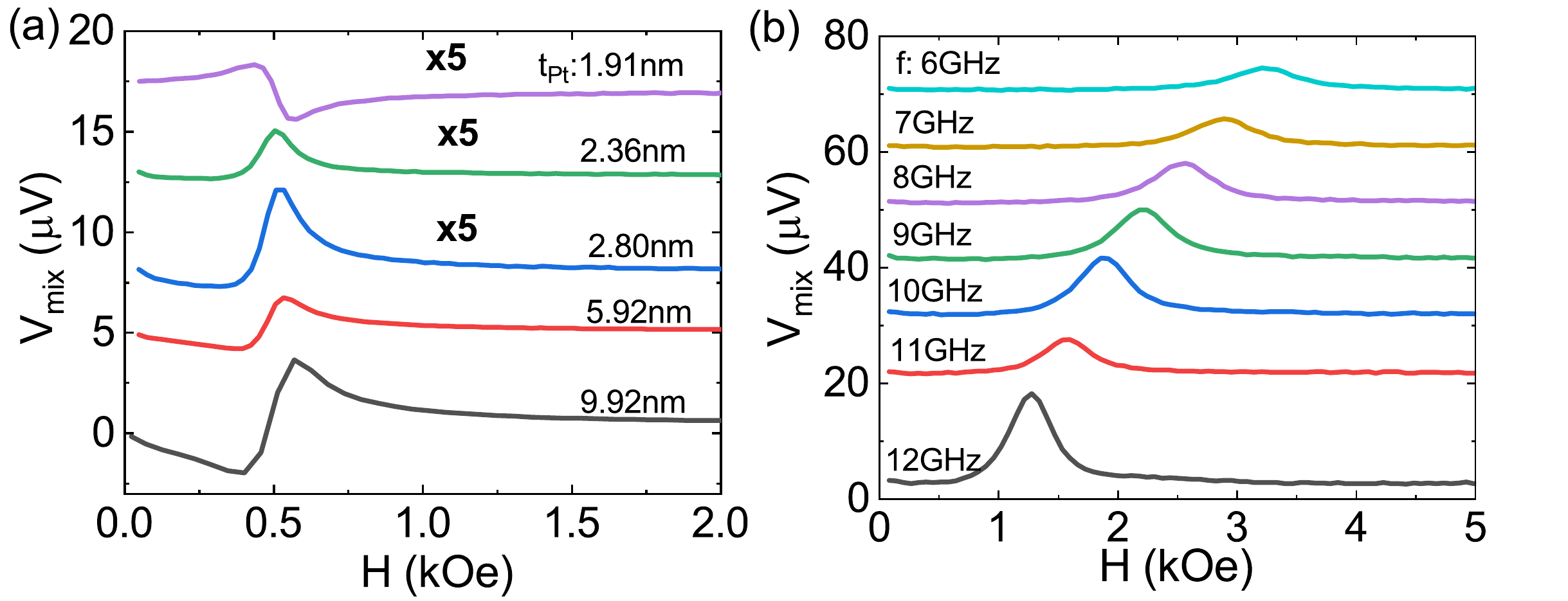}
\caption{$V_{\text{mix}}$ as a function of magnetic field measured in bilayers for different Pt thickness with excitation frequency $f$ = 4 GHz (a) and fixed $t_\mathrm{Pt}$ = 2.36 nm with different $f$ (b). Curves are artificially offset for clarity.}
\label{fig:Vmix}
\end{figure}

Spin transport properties of Pt, such as spin Hall angle and spin diffusion length, were determined using a dedicated model based on Ref.~[\onlinecite{skowronski_2018}]. Mixing voltage, $V_\mathrm{mix}$, was obtained from the SOT-FMR measurement as a function of magnetic field, $H$, and the results are presented in Fig.~\ref{fig:Vmix}. The signal consists of two parts - symmetric and antisymmetric - which are modelled using Lorentz curves, 
\begin{equation}
V_{\text{mix}} = V_S\frac{\Delta H^2}{\Delta H^2+(H-H_0)^2} + V_A \frac{\Delta H(H-H_{0})}{\Delta H^2+(H-H_{0})^2}\,,
\label{eq:Vmix}
\end{equation}
where $V_S$ ($V_A$) is the magnitude of symmetrical (antisymmetrical) Lorentz curve, $\Delta H$ is the linewidth, and $H_0$ is the resonance field.
  
We assume that the greatest contribution to magnetoresistance comes from spin Hall magnetoresistance (SMR) effect which can be described in this case via model developed in Ref.~[\onlinecite{kim_2016}],
\begin{align}
\frac{\Delta R}{R_{\text{Pt}}}&\approx\frac{\theta^2\lambda}{t_{\text{Pt}}}\left[\frac{g_{\text{Pt}}^R}{1+g_{\text{Pt}}^R}-\frac{g_{\text{Pt}}^{\text{Co}}}{1+g_{\text{Pt}}^{\text{Co}}\operatorname{coth}(t_{\text{Pt}}/\lambda)}\right]\times \nonumber \\
&\times\tanh{\frac{t_{\text{Pt}}}{\lambda}}\tanh^2{\frac{t_{\text{Pt}}}{2\lambda}} \nonumber 
\end{align}
where $R_{\text{Pt}}=l\rho_{\text{Pt}}/(t_{\text{Pt}}w)$, and $\lambda$ are the resistance, the spin Hall angle and the spin diffusion length of Pt. Respectively, $g^R_{\text{Pt}}=2e^2/\hbar\lambda\rho_{\text{Pt}}G^R\coth{(t_{\text{Pt}}/\lambda)}$ is the real part of dimensionless spin-mixing conductance with $G^R$ being the real part of spin-mixing conductance, and
\begin{equation}
g_{\text{Pt}}^{\text{Co}}=\frac{(1-p^2)\rho_{\text{Pt}}\lambda}{\rho_{\text{Co}}\lambda_{\text{Co}}\operatorname{coth}(t_{\text{Co}}/\lambda_{\text{Co}})}
\end{equation}
is the dimensionless spin conductivity of Pt/Co interface, where $p$ is the spin polarization of Co at the Fermi level, and $\lambda_{\text{Co}}$ is the spin diffusion length in Co layer~\cite{kim_2016}.



The symmetric component to the signal is known to originate from spin Hall effect induced antidamping-like field\cite{skowronski_2018},
\begin{equation}
H_{\text{DL}} \approx -\frac{\hbar j_c^{\text{Pt}}}{2eM_st_{\text{Co}}}\theta\left[1 - \operatorname{sech}\frac{t_{\text{Pt}}}{\lambda}\right]\frac{g^R_{\text{Pt}}}{1+g^R_{\text{Pt}}}\,,
\label{eq:DL}
\end{equation}
where $M_s$ is the saturation magnetization of Co and $j_{\text{c}}^{\text{Pt}}$ is the current density in Pt layer. Moreover, in the derivation of formula~(\ref{eq:DL}) we have assumed neglible imaginary part of spin-mixing conductance.
The antisymmetric component, on the other hand, is dominated by contributions from Oersted field,
\begin{equation}
H_{\text{Oe}} = -\frac{j_c^{\text{Pt}} t_{\text{Pt}}}{2}\,,
\label{eq:Oe}
\end{equation}
and interfacial Rashba-Edelstein contribution,
\begin{equation}
H_{\text{SO}} = \Gamma_{\text{SO}}\,,
\label{eq:SO}
\end{equation}
which is determined by thickness-independent spin-orbit coupling strength $\Gamma_{\text{SO}}$\cite{skowronski_2018}.
We consider here this contribution as originating from interfacial spin-orbit coupling, and distinct from spin Hall contribution, as predicted theoretically~\cite{haney_2013}, although it is still the subject of ongoing debate. Experimental works, however, seem to confirm the possible Rashba-Edelstein origin of the field~\cite{kawasuso_2013,allen_2015}. Note, that an interfacial contribution might be also attributed to magnetic proximity effect from ferromagnet~\cite{koyama_2017,peterson_2018}, which was recently discussed by Zhu et al~\cite{zhang_2018}. The authors showed that proximity effect in Co/Pt and Au$_{25}$Pt$_{75}$/Co interfaces has minimal correlation on either the dampinglike or the fieldlike spin-orbit torques compared to other interfacial effects like interfacial spin-orbit scattering. The relevant parameters used to modelled of the effective fields are summarized in Table \ref{tab:param}.

\begin{table}[H]
\centering
\caption{Magnetotransport parameters of modelled Co/Pt bilayer system.}
\label{tab:param}
\begin{ruledtabular}
\begin{tabular}{lllll}
Model & Value & Fitted & Value \\  \colrule
$\lambda_{\text{Co}}$ & 1 nm$^\text{a}$ & $\theta$ & 0.08    \\
p & 0.75$^\text{a}$ & $\lambda$ & 2.17 nm    \\ 
$G^R/(e^2/ \hbar)$ & 3.9$\times 10^{19}$ $\Omega^{-1}\text{cm}^{-2}$       & $\Gamma_{\text{SO}}$ & 1.5 Oe \\ \colrule
\end{tabular}
$^\text{a}$Ref.~[\onlinecite{kim_2016}]
\end{ruledtabular}
\end{table}

\begin{figure}[t!]
\centering
\includegraphics[width=\columnwidth]{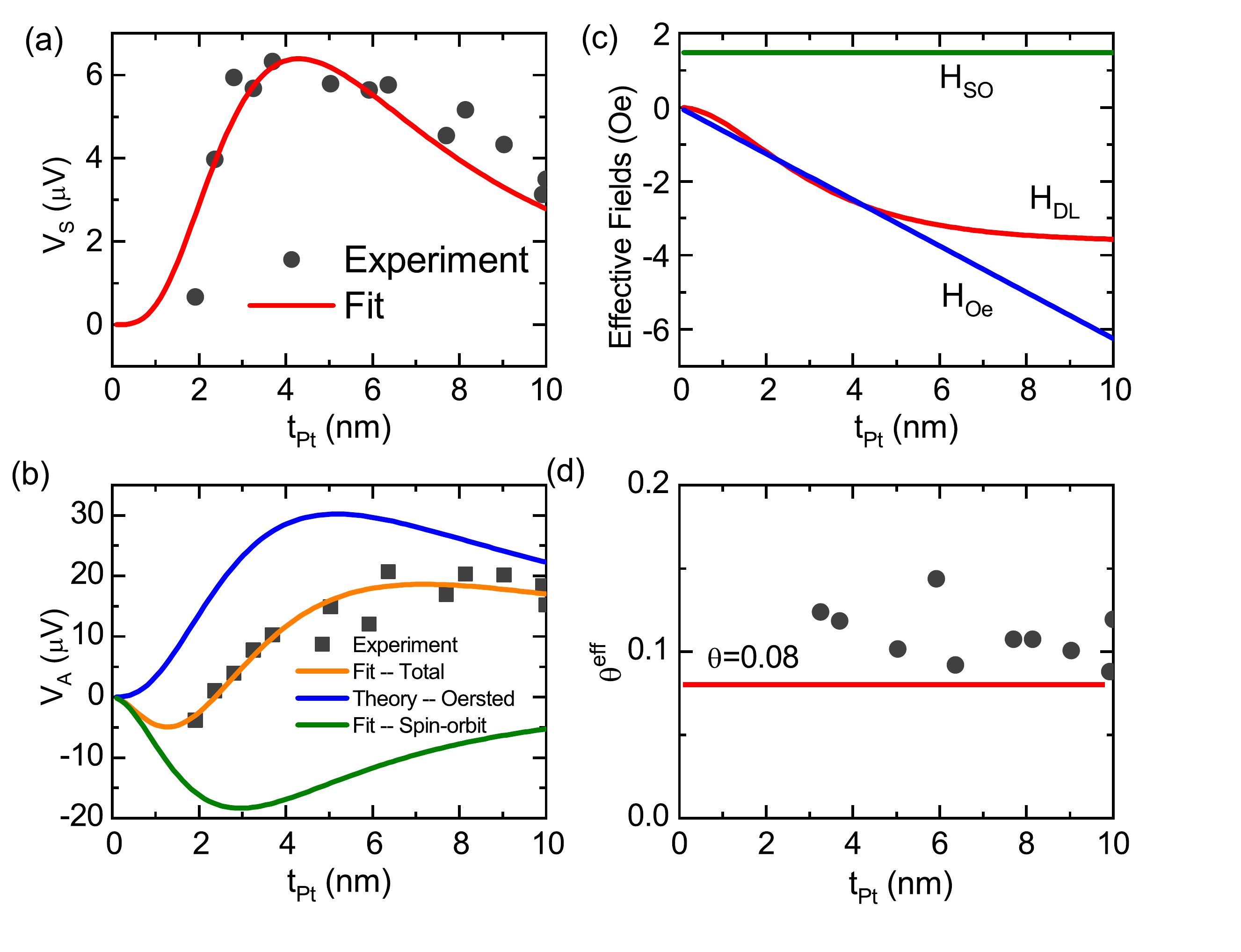}
\caption{The experimental and fitted from Eqs.~(\ref{eq:DL})-(\ref{eq:SO}) amplitudes of symmetric, $V_\text{S}$, and antisymmetric, $V_\text{A}$, components to the signal, $V_{\text{mix}}$, are shown as functions of thickness $t_{\text{Pt}}$ in (a) and (b), respectively. The contribution of each effective field is shown in (c). Effective spin Hall angle and spin Hall angle of Pt obtained from fitting Eq.~(\ref{eq:DL}) to the data are shown in (d).}
\label{fig:Model}
\end{figure}


The components $V_S$ and $V_A$, obtained from Lorentz curve modelling for Co/Pt bilayers, are presented in Fig.~\ref{fig:Model}(a-b) as functions of $t_\mathrm{Pt}$. These dependencies are further fitted with Eqs.~(\ref{eq:DL})-(\ref{eq:SO}). As a result, the spin Hall angle of $\theta = 0.08$ is obtained, which agrees with literature values for this material~\cite{skowronski_2018,sanchez_2014}. Note, that in Fig.~\ref{fig:Model}(d) the effective spin Hall angle, $\theta^{\text{eff}}\approx 0.1$, obtained roughly from ratio $V_S$/$V_A$ is slightly larger than the spin Hall angle of Pt, $\theta=0.08$, obtained from fitting the model derived in Ref.~[\onlinecite{skowronski_2018}] to the data. As shown there, the spin Hall angle is treated as a bulk property of heavy metal layer and agrees with effective spin Hall angle for thick Pt.

\begin{figure}[H]
\centering
\includegraphics[width=\columnwidth]{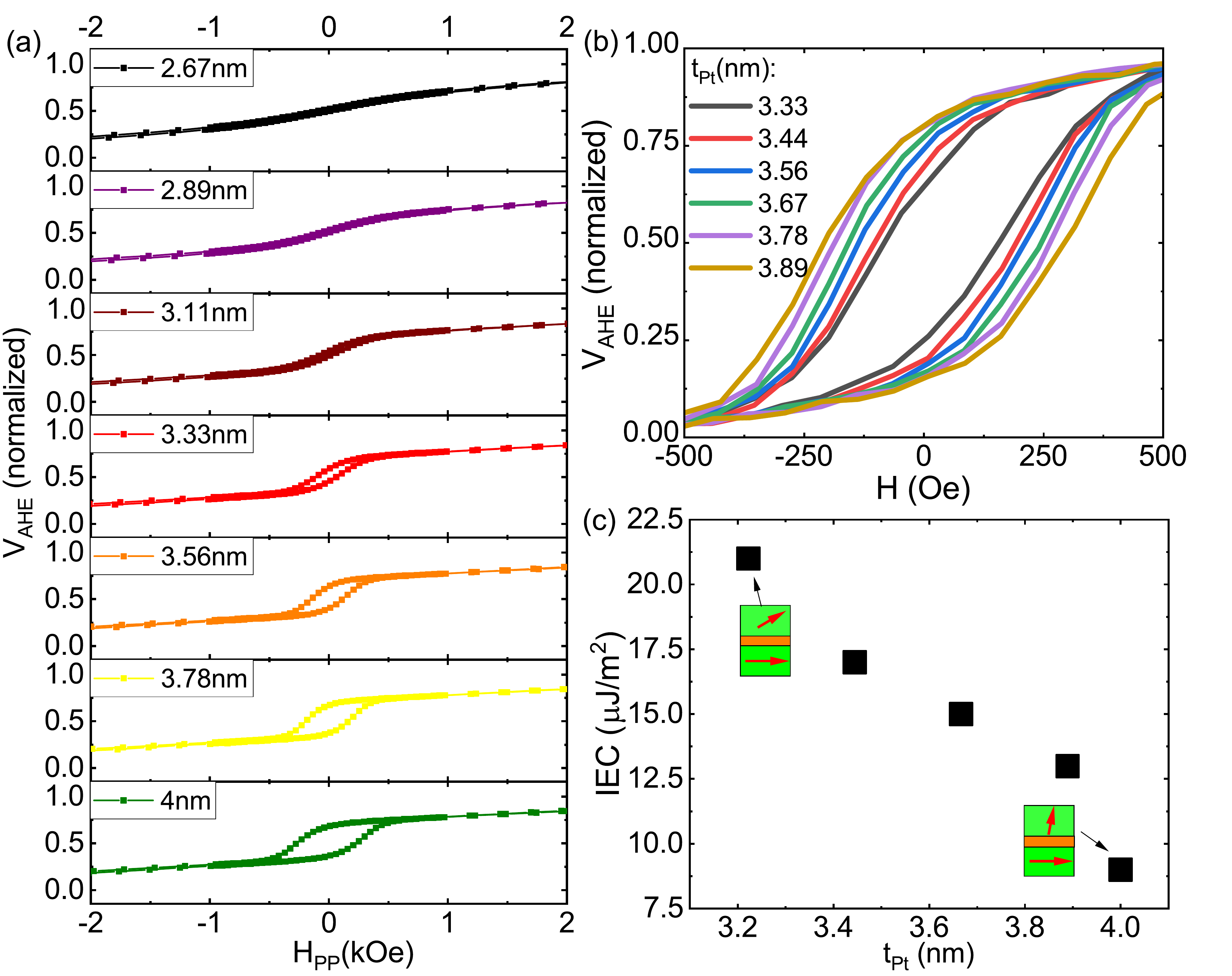}
\caption{AHE hysteresis loops measured in trilayers with different Pt spacer thickness (a). Open hysterisis loops are measured for $t_\mathrm{Pt}$ > 3 nm  - (b). IEC energy as a function of $t_\mathrm{Pt}$ is depicted in (c). Cartoon presents effective magnetization direction in top and bottom Co layers.}
\label{fig:AHE}
\end{figure}

Next, we focus on multilayer structures with two Co layers (3 nm and 1 nm) separated by Pt spacer of different thickness. Thicker (thinner) Co is characterized by in-plane (perpendicular) magnetic anisotropy, which is manifested by the shape of the anomalous Hall effect (AHE) hysteresis loop (Fig.~\ref{fig:AHE}(a)). In addition, these two layers are ferromagnetically coupled, and the energy of this coupling varies with $t_{Pt}$.\cite{matczak_2015} To quantitatively determine the coupling energy, AHE curves for different Pt spacer were modelled using macrospin approach\cite{czapkiewicz_2004}, and the energy values are depicted in Fig.~\ref{fig:AHE}(c). For high coupling energies (corresponding to $t_\mathrm{Pt}$ < 3 nm) the AHE loop is hysteresis-free. For the spacer thickness of $t_\mathrm{Pt}$ > 3 nm hysteresis in AHE is observed, which enables SOT-induced switching presented below.


\begin{figure}[H]
\includegraphics[width=\columnwidth]{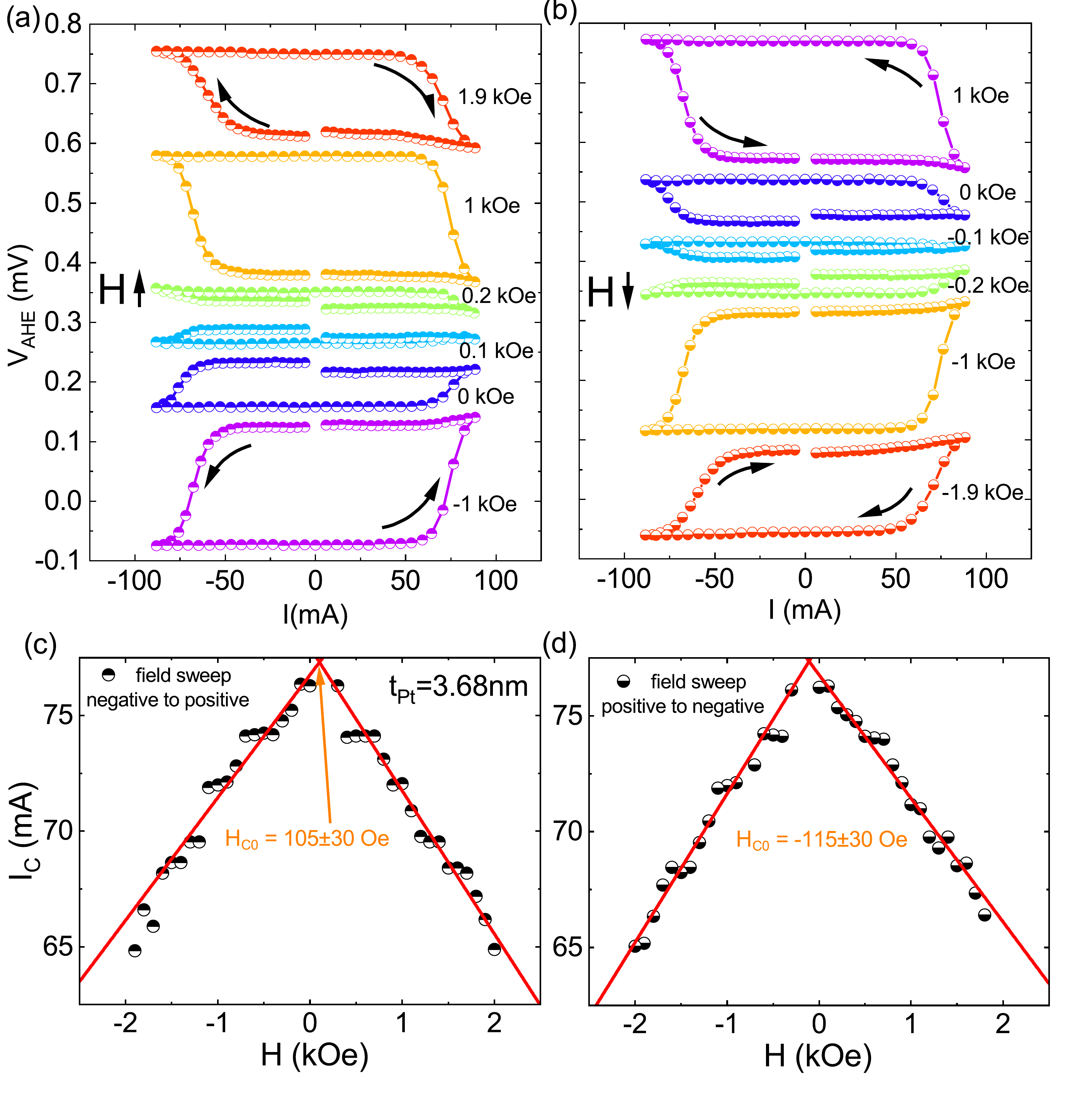}
\caption{SOT-induced switching process under different external magnetic fields for trilayer with a spacer thickness of $t_\mathrm{Pt}$ = 3.68 nm. Hall voltage hysteresis loops as a function of in-plane current measured under various in-plane magnetic fields changed from negative to positive (a) and positive to negative (b) values. Corresponding critical currents for different fields are presented in (c) and (d).}
\label{fig:switching_fields}
\end{figure}

As shown above, the top Co layer is tilted from the perpendicular direction due to ferromagnetic coupling with a bottom in-plane magnetized Co layer. Such symmetry breaking is essential for a field-free SOT-induced magnetization switching. Figure \ref{fig:switching_fields}(a-b) presents magnetization state of the top Co (described with AHE voltage) as a function of in-plane current ($I$) for different in-plane magnetic fields, when the magnetic field is increased (a) and decreased (b). Clockwise and counter-clockwise loops are observed, which depends on the tilt direction with respect to the surface of the top Co layer magnetization. 
The switching current depends linearly on magnetic field value according to Ref.~[\onlinecite{lee_2013}]. However, the transition from counter-clockwise to clockwise loops occurs at around 150 Oe for increasing field and around -150 Oe for decreasing field. This change in transition originates from the in-plane magnetized bottom Co switching, and is different than in the case of heterostructures using antiferromagnets~\cite{fukami_2016}. 

The critical currents and current densities for several Pt thicknesses are summarized in Tab.~\ref{tab:currents}. As expected, critical current needed to switch top Co magnetization without external magnetic field decreases with increasing IEC energy.

\begin{table}[t!]
\centering
\caption{Summary of critical switching currents, $I_{\text{c0}}$, and current densities, $J_{\text{c0}}$, of Co/Pt/Co trilayers for different Pt thickness, $t_{\text{Pt}}$.}
\label{tab:currents}
\begin{ruledtabular}
\begin{tabular}{ccc}
$t_{\text{Pt}} (\text{nm})$ & $I_{\text{c0}} (\text{mA})$ & $J_{\text{c0}} (\text{A}/\text{m}^2)$\\ \colrule
3.7 & 77 & $7.00\cdot 10^{11}$               \\ 
3.6 & 70 & $6.50\cdot 10^{11}$              \\ 
3.4 & 61 & $6.02\cdot 10^{11}$             \\
2.9 & 51 & $5.95\cdot 10^{11}$           \\ \colrule
\end{tabular}
\end{ruledtabular}
\end{table}

\begin{figure}[b!]
\includegraphics[width=\columnwidth]{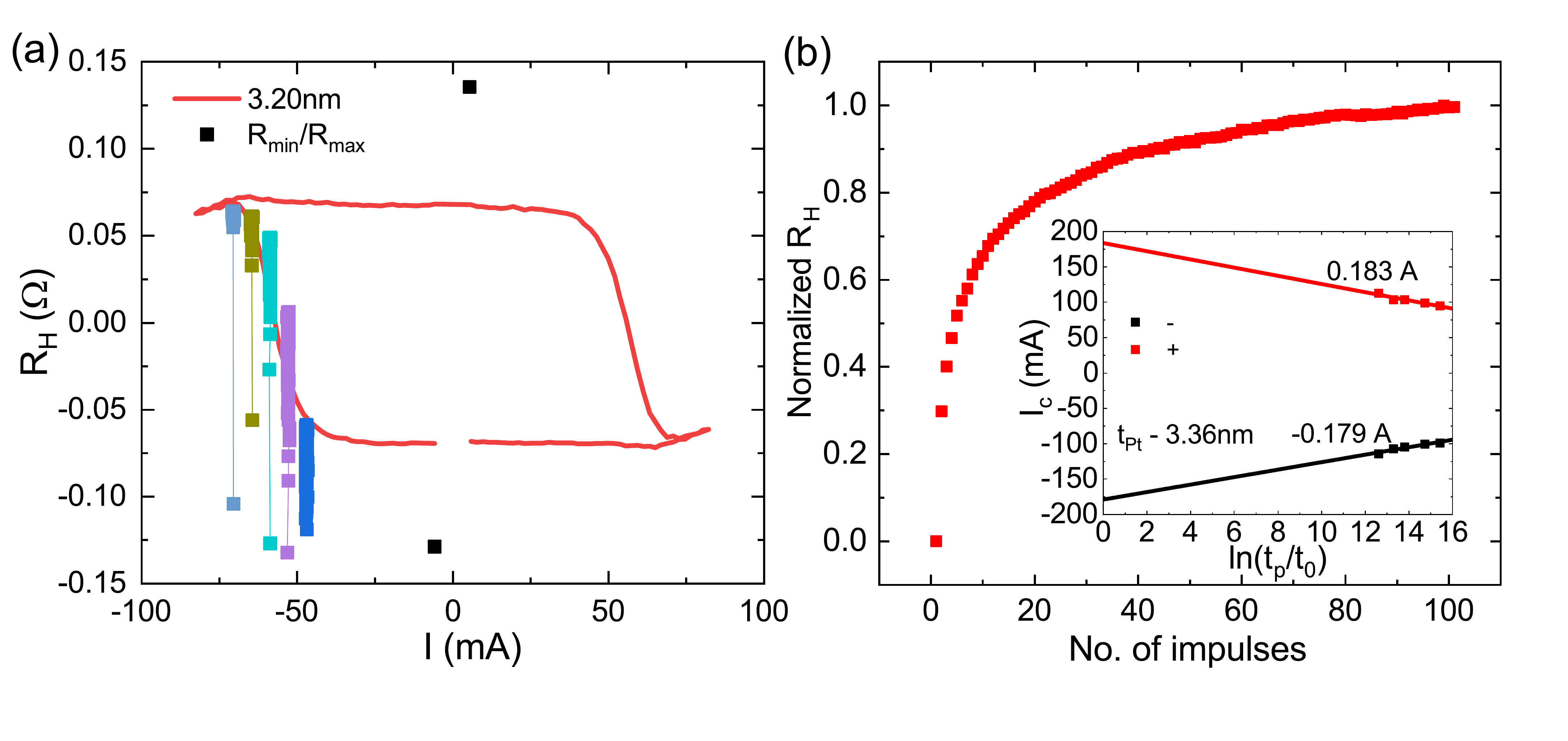}
\caption{Hall resistance loops as a function of current for $t_\mathrm{Pt}$ = 3.2 nm. Full black squares in (a) corresponds to the initial resistance after saturation in a magnetic field. Squares of different colours indicate the resistance state after application of up to a hundred current pulses of given amplitude. Normalized resistance after different number of pulses of around 60 mA amplitude is presented in (b). Both parts (a) and (b) highlights the neuromorphic-like behavior of the sample. Inset: the switching current density dependence on the switching pulse duration ($t_p$). $t_0$ is the inverse of the attempt frequency equal to 1 ns.}
\label{fig:switching_train}
\end{figure}

Finally, we move on to the description of the analogue SOT-switching of thin Co layer. As seen in SOT-switching experiments, AHE voltage changes gradually with current, contrary to current-induced switching required typically for digital memory applications ~\cite{liu_2012}. Figure \ref{fig:switching_train}(a) presents Hall resistance $R_\mathrm{H} = V_{\text{AHE}}/I$ of a trilayer with a Pt spacer thickness of $t_\mathrm{Pt}$ = 3.2 nm.

 After initialization in a saturating perpendicular magnetic field, $R_\mathrm{H}$ changes from a minimal value of $R_\mathrm{H-min}$ =  -0.14 Ohm to around 0.07 Ohm upon negative current application. Note that maximal resistance for this spacer thickness is $R_\mathrm{H-max}$ = 0.14 Ohm, so no saturation is achieved during SOT-switching. Sweeping the current from negative to positive values induces switching to a lower $R_\mathrm{H}$; however, only down to around $R_\mathrm{H}$ = -0.07 Ohm, which is above $R_\mathrm{H-min}$. Therefore, during SOT-induced switching, the magnetization state is not reversed fully, but possibly magnetization domain state is formed\cite{baumgartner_2017}. Moreover, $R_\mathrm{H}$ depends also on the number of pulses of the same amplitude, as well as the pulse duration - Fig. \ref{fig:switching_train}(b). By decreasing the switching pulse duration time ($t_p$) from 500 ms down to 500 $\mu$s, the absolute value of the critical current increases according to Ref.~[\onlinecite{kubota_2006}]. Nonetheless, such gradual change in Hall resistivity mimics behaviour of a neuron, whose resistive potential changes based on electrical signals delivered by synapses in an analogue-manner ~\cite{krzysteczko_2009}.


\section{Summary}
In summary, SOT-induced switching in Co/Pt/Co structure with in-plane and perpendicular anisotropy of Co layers coupled ferromagnetically by Pt spacer with varying thickness was investigated. 
The spin Hall angle and spin diffusion length were determined using SOT-FMR method, the results of which were analyzed using dedicated theoretical model. 
The coupling between two Co layers is tuned by the Pt spacer thickness and for intermediate thickness, $2.9~\text{nm}<t_{\text{Pt}}<3.7~\text{nm}$, field-free SOT-induced switching of perpendicularly magnetized Co layer was observed. For a range of Pt spacer thickness, gradual magnetization change with in-plane current was obtained, which may be useful for a hardware implementation of a spintronic neuromorphic network. 
\section*{Acknowledgments}
This work is supported by the National Science Centre, grant No. UMO-2015/17/D/ST3/00500, Poland. S. Z. acknowledges grant No. LIDER/467/L-6/14/NCBR/2015 by the Polish National Centre for Research and Development. S. \L{}., \L{}. K., K.G. and T. S. acknowledge National Science Centre grant Spinorbitronics UMO-2016/23/B/ST3/01430. Nanofabrication was performed at the Academic Center for Materials and Nanotechnology of AGH University of Science and Technology.

\end{document}